\documentclass[a4paper,11pt]{article}
\pdfoutput=1 % if your are submitting a pdflatex (i.e. if you have
\usepackage{jcappub} % for details on the use of the package, please
\usepackage{graphicx}

\newcommand{\EGMF}{Extragalactic Magnetic Field}

\newcommand{\CRPropa}{CRPropa 3 \cite{batista_crpropa_2016}}

\title{The 320 EeV Fly's Eye event: a key messenger or a statistical oddity?}
\author{Thomas Fitoussi,}
\author{Gustavo Medina-Tanco,}
\author{Juan Carlos D'Olivo}
\affiliation{%
    Instituto de Ciencias Nucleares, Universidad Nacional Aut\'onoma de M\'exico, 
    Circuito Exterior,  C.U., 04510, CDMX, M\'exico
}%

\emailAdd{thomas.fitoussi@correo.nucleares.unam.mx}
\emailAdd{gmtanco@nucleares.unam.mx}
\emailAdd{dolivo@nucleares.unam.mx}

\arxivnumber{1906.11170}

\abstract{
Almost three decades ago, the Fly's Eye experiment recorded the most energetic cosmic-ray ever observed. With an energy of 320 EeV, this event is well beyond the suppression region of the ultra-high energy cosmic rays (UHECR) spectrum. Modern and larger observatories, with an exposure up to 60 times larger, have never observed an event with even remotely comparable energy. Thus, if the energy of the Fly's Eye event was indeed well measured, as strongly suggested by the data, then it remains a great mystery or an unbelievable stroke of luck. At such high energies, the Universe is very opaque to electromagnetic interacting particles, whether photons, protons or heavy nuclei, and therefore its source must be relatively close. Using numerical simulations for the propagation of protons and nuclei, we reexamine the problem by testing different possibilities for the nature of the primary, the injection spectrum and the location of the source. Based on these calculations, we show that the most feasible scenario corresponds to a nearby ($\sim 2-3$ Mpc) bursting source of heavy nuclei in the northern sky, which injected a hard spectrum ($\gamma \le 1.5$) with an energy cut-off between $300$ and $1000$ EeV. Such scheme generates a significant probability for the observation of one event by Fly's Eye combined with a null result of Telescope Array at the same energy. 
}


\begin{document}
\maketitle
\flushbottom

\section{Introduction}

On October 15, 1991, the Fly's Eye (FE) experiment in Utah recorded the most energetic cosmic-ray ever observed. The so called Fly's Eye event has a very well determined energy of $320^{+92}_{-94}$ EeV with a clean fluorescence profile. Moreover, the sky was clear that night and the atmospheric cascade created by this event did not point towards the detector, which minimizes the uncertainty due to contamination by direct Cherenkov light. Therefore, the estimate of 220 EeV seems to be a rather solid lower limit for the energy of the event, putting it far beyond the GZK cut-off \citep{bird_detection_1995}. Indeed, protons with such a high energy cannot travel more than few tens of Mpc due to photo-pion production interaction with the CMB \citep{greisen_end_1966,zatsepin_upper_1966}. Heavier nuclei, on the other hand, are effectively absorbed on an even shorter distance because of photo-disintegration \citep{allard_extragalactic_2012}. 

The nature of this cosmic ray with extremely high energy has not been properly identify. The first study of the event \citep{bird_detection_1995} was not able to determine the exact nature of the primary. More recently,  the depth of the shower maximum $X_{max}$ and the longitudinal profile of the event were recomputed, with equally inconclusive results \citep{risse_primary_2004,risse_primary_2006}. The primary cosmic-ray can be either a proton or a heavier nuclei (C, Fe). In fact, even a photon event cannot be completely ruled-out at more than two sigmas. The arrival direction of the FE event is instead reasonably well established at $\sim 20^\circ$ from the anti-galactic center (right ascension: $85.2^\circ \pm 0.48^\circ$, declination: $48^\circ \pm 6^\circ$, galactic latitude: $9.6^\circ$ , galactic longitude: $163.4^\circ$ \cite{bird_detection_1995}). On the other hand, the location of the source on the sky is poorly constrained, due to the deflections of the particle by the intergalactic magnetic fields which are not well known \cite{durrer_cosmological_2013}. From the probability of absorption of a cosmic ray, it has been shown that the source can not be further away than few Mpc for heavy nuclei and a hundred of Mpc for protons \citep{sigl_origin_1994,elbert_search_1995}, or not more than tens of Mpc at 95\% CL, according to our own calculations (see below). 

Unless the event was a statistical fluke and the cosmic ray travelled from the source to the Earth without interacting with the photon background, the primary particle must have left the source at an even higher energy. This make the observation of the air shower produced by it more intriguing in the absence of a counterpart in detectors with a much larger exposure. Even more, non obvious source in other high energy channels, e.g., X-rays, gamma-rays or neutrinos, is recognizable in the region of the sky around the direction of the FE event, which could be plausibly associated with it. Much the same can be said on radio galaxies, which are promising candidates, and of which there are not many inside a 100 Mpc. Nonetheless, some candidates were considered in the literature, including quasars and AGNs \citep{sigl_origin_1994,rachen_possible_1995}, as well as a strong gamma-rays burst observed by BATSE near the location of the event \citep{milgrom_possible_1995}. In a more recent study \citep{gnatyk_search_2016}, the particle is backtracked for a specific realization of the Galactic and extragalactic magnetic fields, with cataclysmic events (like the birth of a millisecond pulsar or a magnetar flare) in galaxies UGC04874 or UGC03394 suggested as possible sources for a C, N, O or Fe primary. 

To the extent that its nature and origin remain unknown, the FE event continues to be challenging for physics and astrophysics after almost three decades of its observation. Particularly puzzling is the fact that no other event of such energy has been detected since then, even by subsequent larger and more sensitive detectors: HiRes, Pierre Auger Observatory (PAO) or TA. The total exposure of FE was of only 820 km$^2$ sr yr \citep{bird_cosmic-ray_1994,bird_results_1995}, while the one of HiRes was of $\sim$ 5000 km$^2$ sr yr \citep{bergman_cosmic_2007,bergman_observation_2007,abbasi_first_2008,belz_overview_2009}, 
and those of TA and PAO are of 8300 and 67000 km$^2$ sr yr \citep{ivanov_energy_2012,the_telescope_array_collaboration_pierre_2018}, respectively. The UHECR sky in the South may differ from the one in the North, as indeed the data of PAO and TA seem to indicate \citep{PAOandTA_2018},  rendering  the potential source invisible  to the PAO. However, the problem still persists for HiRes and TA, which are at the same location as FE and have not seen an event with such high energy. Also intriguing is the fact that FE never observed a single event at 100 EeV despite the reported detection at 320 EeV. 

Having said that, we have not been able to find in the literature any strong argument against the physical reality of the FE event. Therefore, it may be worth having a new look at the problem, examining the implications of this event in the context of the scenario that has been developing in the last years as a result  of the measurements of the largest experiments currently in operation.

The organization of the paper is as follows.
In Sec. \ref{sec:simple_model}, we explore with the aid of a simple model the likelihood of the FE event and the simultaneous null results of TA and HiRes, assuming that the true UHECR background spectrum is well described by TA. We extend this analysis by adding a secondary component and check the impact of the normalization and the hardness of the spectrum on the before mentioned probability. 
In Sec. \ref{sec:iron} and \ref{sec:proton}, we use numerical simulations to analyze the impact of the propagation of iron nuclei and protons on the observation of the secondary component. Then, we use that information to propose a plausible scenario. Finally, our conclusion are presented in Sec. \ref{sec:conclusion}.

\section{Exploratory model}
\label{sec:simple_model}

Some preliminary insight into the problem can be gained by using a simple model, in which the effects of propagation are neglected. 

For the purpose of the paper, we assume that the UHECR spectra published by each experiment, regarding shape, is a good estimator of the incoming flux. Nevertheless,  we apply an energy rescaling to account for systematics uncertainties and bring them into agreement in the energy region between $10^{19}$ and $10^{19.3}$ eV. We start by applying the rescaling done in the joint analysis lead by PAO and TA (-5.2\% for TA). The FE spectrum is then rescaled by a factor of -20\%, which is compatible with the published systematic errors \citep{bird_results_1995}, while the HiRes spectrum is rescaled by -5\%.

\begin{figure}[hbtp] \centering
    \includegraphics[width=0.8\textwidth]{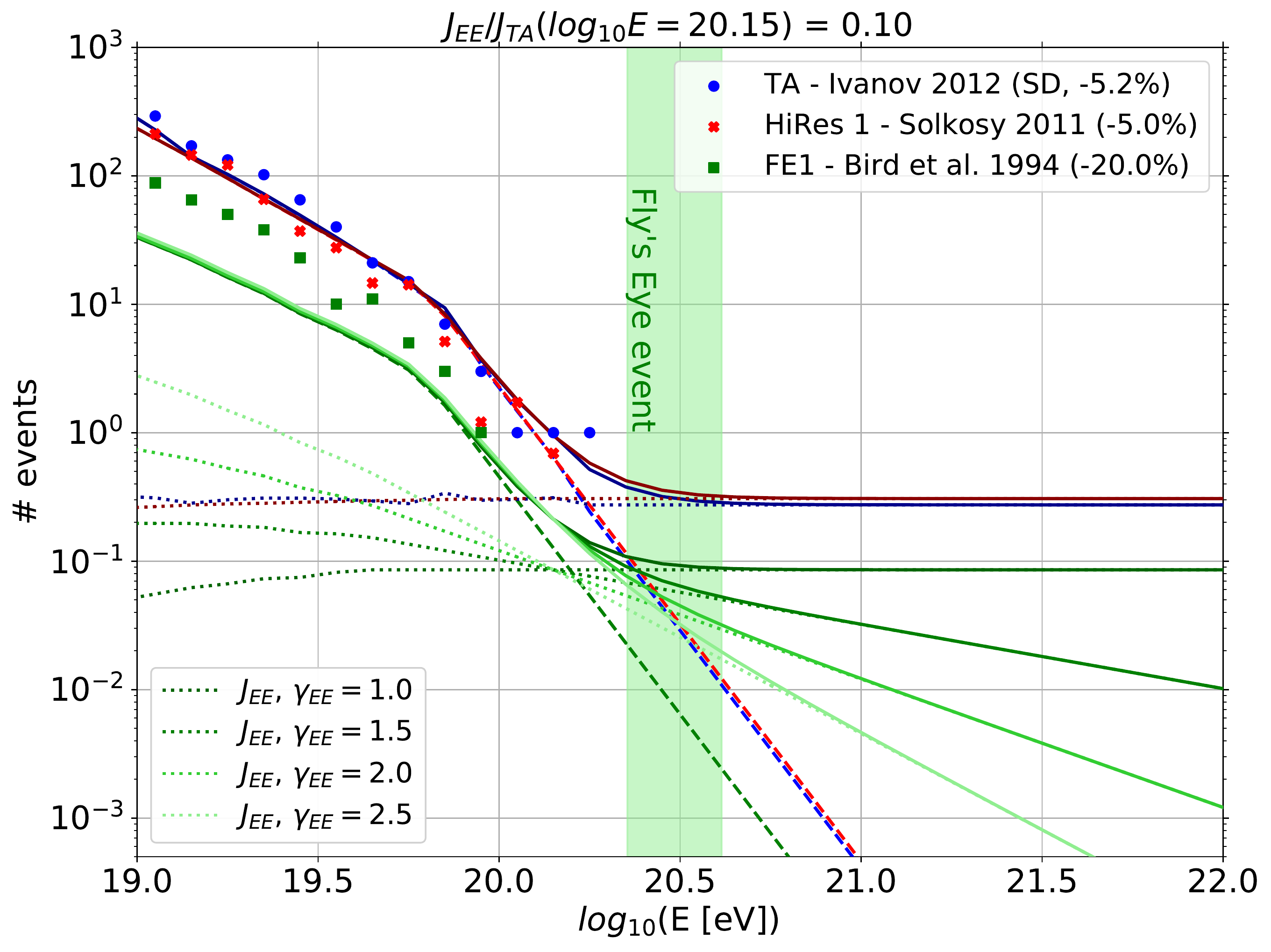}
    \caption{Number of events expected for the exposures of TA, HiRes and FE (blue, red and green dashed lines, respectively), by assuming that the northern sky UHECR flux is well described by the TA spectrum. A hypothetical secondary component is depicted in dotted lines for different power law indexes $\gamma_{EE}$:  $\gamma_{EE}=1$ for HiRes (blue) and TA (red), $\gamma_{EE}=1$, 1.5, 2, 2.5 for FE (green from dark to light). The combined spectra are depicted with solid lines (same colors).}
    \label{fig:Nevts_by_hand}
\end{figure}

Fig. \ref{fig:Nevts_by_hand} shows the number of events as a function of energy for each experiment, computed by multiplying the corresponding rescaled spectra by its exposure in each energy bin \citep[data taken from][]{bird_cosmic-ray_1994,sokolsky_observation_2009,ivanov_energy_2012}. Blue dots are for TA surface detectors, red crosses for HiRes monocular and green squares for FE monocular. 

Consider the all-sky spectrum as fitted by TA \citep{verzi_measurement_2017}:
\begin{equation}
J_{TA}(E)\propto
\left\{\begin{array}{rcl}
E^{-\gamma_1}\,,& &E < E_{\rm ankle}\,,  \\
E^{-\gamma_2}\,,& & E_{\rm ankle} < E < E_{\rm break}\,, \\
E^{-\gamma_3}\,,& & E > E_{\rm break}\,, \label{eq:J_TA}
\end{array} \right.
\end{equation}
\noindent where $E_{\rm ankle} = 5.2 \pm 0.2$ EeV, $E_{\rm break} = 60 \pm 7$ EeV, $\gamma_1 = 3.226 \pm 0.007$, $\gamma_2 = 2.66 \pm 0.02$ and $\gamma_3 = 4.7 \pm 0.6$. 

Assuming that, to a considerable degree, TA, HiRes and FE see the same sky and that the energy spectrum is well described by TA, using Eq. \ref{eq:J_TA} we compute the expected number of events for the three observatories. The results, normalized according to the respective exposures, are presented in  Fig. \ref{fig:Nevts_by_hand} (TA in blue, HiRes in red, and FE in green). Let us denote by $P(\geq1)$ the probability of detecting at least one particle. If such a component were the only one  present in the UHECR flux and could be extrapolated to higher energies with the same power law index, then we see that $P(\geq1) \sim 5 \times 10^{-3}$, which means that the occurrence of the FE event would be a hundred times beyond expectations. Even more, although TA has an exposure an order of magnitude greater, so far it has not detected a single event in the same spectral region. These facts highlight the peculiarity of the FE event, seeming to point towards a statistical fluke. One can envision at least two kinds of outliers: either a particle that traveled from a distant source several mean free paths without, or with very few, interactions with the background photons, or a nearby source that is responsible for increasing the probability of FE observing an event of 300 EeV and, at the same time, allows a non-negligible probability that TA has not seen anything comparable.

\begin{figure}[hbtp] \centering
    \includegraphics[width=0.8\textwidth]{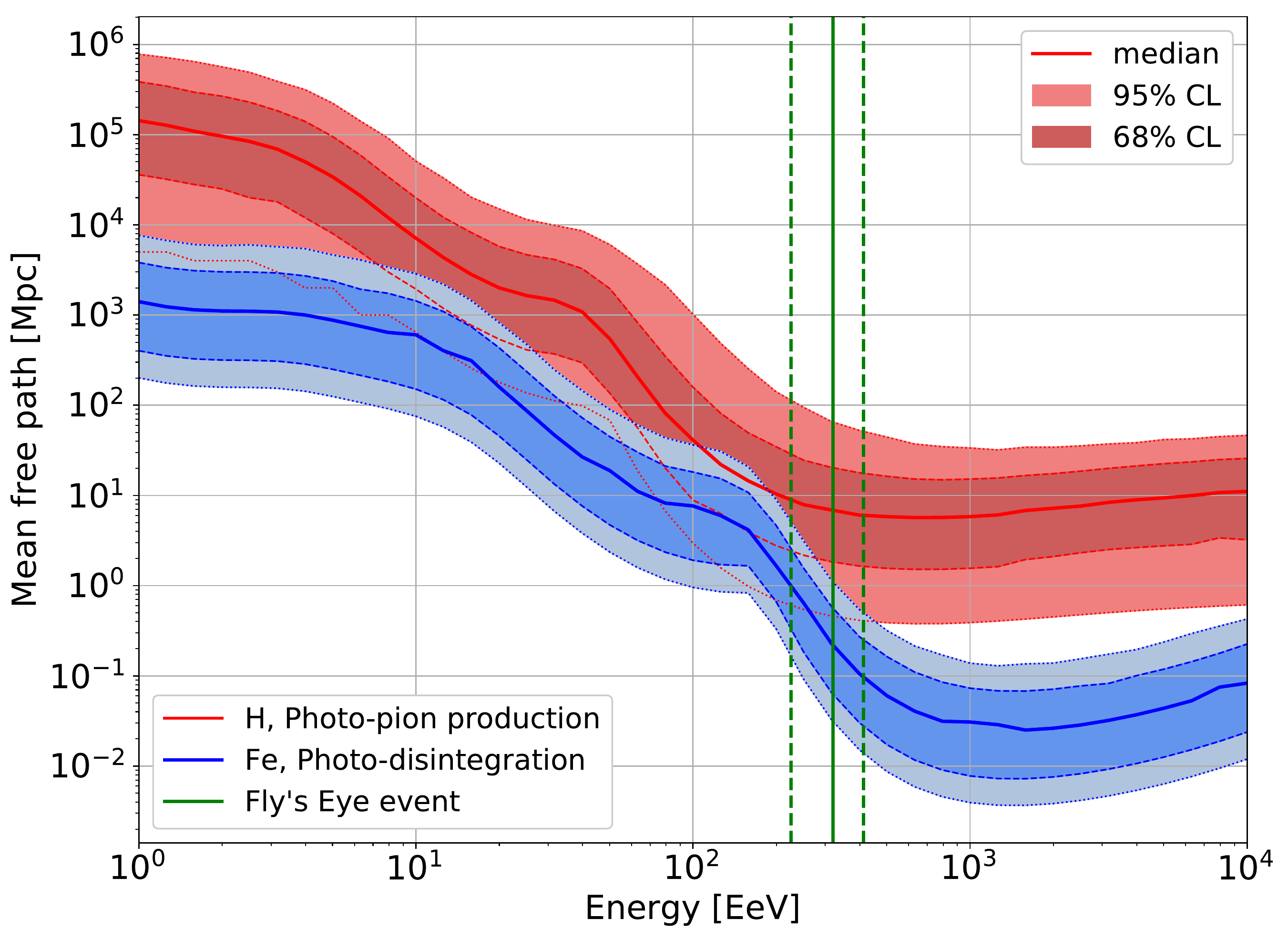}
    \caption{Mean free path of protons and Fe nuclei due to photo-pion production and photo-disintegration, respectively, as a function of energy due to interaction with the photon backgrounds involved.}
    \label{fig:Fe_H_meanFreePath}
\end{figure}

Although the first scenario can never be completely ruled out, as Fig. \ref{fig:Fe_H_meanFreePath} shows, for protons the mean free path increases rapidly as the energy decreases and it is therefore difficult to understand how, for a given distance to the source, particles of lower energy have not being detected.  In the event that the primary was an iron nuclei, the mean free path is so small that the source cannot be far away; indeed, it should probably be inside or near the Local Group. Besides, Fig. \ref{fig:Fe_AvsD}a indicates that, regardless of the energy injected at the source, the energy of the leading fragment degrades below the energy range of the FE event after propagating a few Mpc. In the case of a nearby source, it should be transient, presumably bursting, so that its activity in other messengers, like high energy X-ray or $\gamma$-ray photons, declined to undetectability long ago and only charged particles are arriving at present, delayed by their deflection in the magnetic fields involved. 

In order to roughly model the second scenario, we introduce into the spectrum an additional harder component, which would be associated to a rather local source within the region of the sky that corresponds to the field of vision common to both experiments, FE and TA. The flux of this secondary component extends to extreme energies (EE) and is featured by a power law with an spectral index 
$\gamma_{EE}$:
\begin{equation}
    J_{EE}(E) \propto E^{-\gamma_{EE}}.
\end{equation}
The normalization is done in such a way that TA observations are not violated, namely, requiring that  $J_{TA} + J_{EE}$ does not exceed the flux inferred by TA at $10^{20.15}$ eV and does not predict more than one event at higher energies. Accordingly, we impose  the condition  $J_{EE} / J_{TA} (\log E = 20) \leq 0.1$. 
 
\begin{figure}[hbtp] \center
    \includegraphics[width=0.8\textwidth]{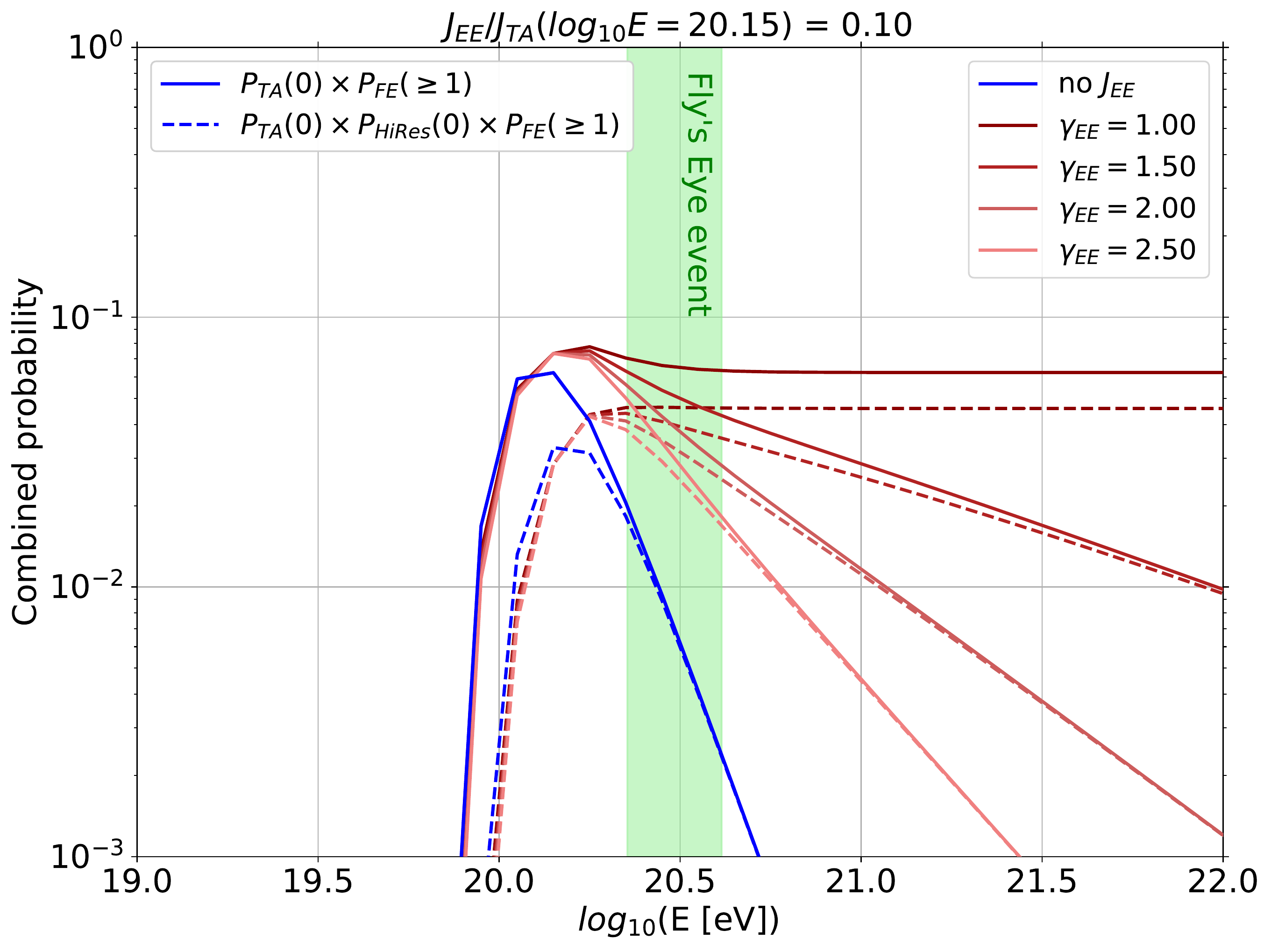}
    \caption{Poissonian probability that FE see at least one event ($P_{FE}(\geq1)$) while HiRes and TA see none ($P_{HiRes}(0)$ and $P_{TA}(0)$) versus the energy, considering that they see the same spectrum ($J_{TA} + J_{EE}$). A power law spectrum with spectral index $\gamma_{EE} = 1$, 1.5, 2, 2.5 is injected by the source.}
    \label{fig:proba_by_hand}
\end{figure}

The dotted lines in Fig. \ref{fig:Nevts_by_hand} show the number of expected events for FE due to the secondary component for different hardness of the spectrum: $\gamma_{EE} = 1$, 1.5, 2, 2.5. The blue and red dotted lines represent this component for $\gamma_{EE} = 1$, normalized to HiRes and TA, respectively. The combined contributions of both components are plotted using solid lines.
From each predicted number of events, we can estimate the Poissonian probabilities $P_{TA}(0)$ ($P_{HiRes}(0)$) of TA (HiRes) detecting zero events and $P_{FE}(\geq1)$ of FE detecting at least one event. Since these measurements are independent, the probabilities for the combined observations are given by the product of the individual probabilities. They are shown in Fig. \ref{fig:proba_by_hand} for different values of $\gamma_{EE}$. The solid lines corresponds to the combined probabilities for TA and FE, while those for the observations of the three detectors, TA, FE and HiRes, are plotted in dashed lines.

As seen from Fig. \ref{fig:proba_by_hand}, without a secondary component (blue lines), the combined probability has a single strong peak at $\sim 10^{20.3}$ eV, but it flattens at energies above this value when a sufficiently hard secondary spectrum ($1 \leq \gamma_{EE} < 1.5$) is incorporated. Therefore, such additional component seems to be a necessary ingredient to bring the joint probability of the results of all experiments to a level compatible with their observations. It should be noted that, although small, the resulting probability is not negligible. 
The simple model considered here does not include the propagation of the particles between the source and the observer and, therefore, the conclusions are mainly qualitative. In what follows, we use numerical simulations to analyze the impact that propagation have on the observation of the secondary component. First we consider that the primary particle is an iron nuclei and the we examine the case of a proton.

\section{FE event as an iron nuclei}
\label{sec:iron}

As already mentioned, in order to properly interpret observations on Earth, it is important to assess the effects of the UHECR propagation  in the intergalactic medium, which are largely dependent on the nature of the injected particle.
The composition of cosmic rays at the highest energies is still an open question. Above $10$ EeV, the PAO claims a decrease in the elongation rate accompanied by a decrease in its shower-to-shower fluctuations. Both effects suggest a gradual increase in the average mass number up to $A \sim 14 - 20$ at $E \sim 30$ EeV. What is the composition at even higher energies is anybody's guess. Although initial TA data seemed to favor a lighter, even pure proton, composition  \citep{abbasi_indications_2010}, the  Auger-TA joint working group has recently concluded that the results of the two experiments are compatible at all energies once systematic and statistical uncertainties have been taken into account \citep{PAO_TA_comp_2016, Hanlon_comp_2018}. To better understand the influence that the identity of the initial particle may have, we will analyse the extreme cases of an iron nucleus ($^{56}$Fe) and a proton. First, we will examine tthe possibility that the FE event originated as an iron nucleus and lost energy and nucleons through photo-disintegration during its journey to the Earth. The case in which the emitted particle could have been a proton will be considered in the next section.

\begin{figure}[hbtp] \center
    \includegraphics[width=0.8\textwidth]{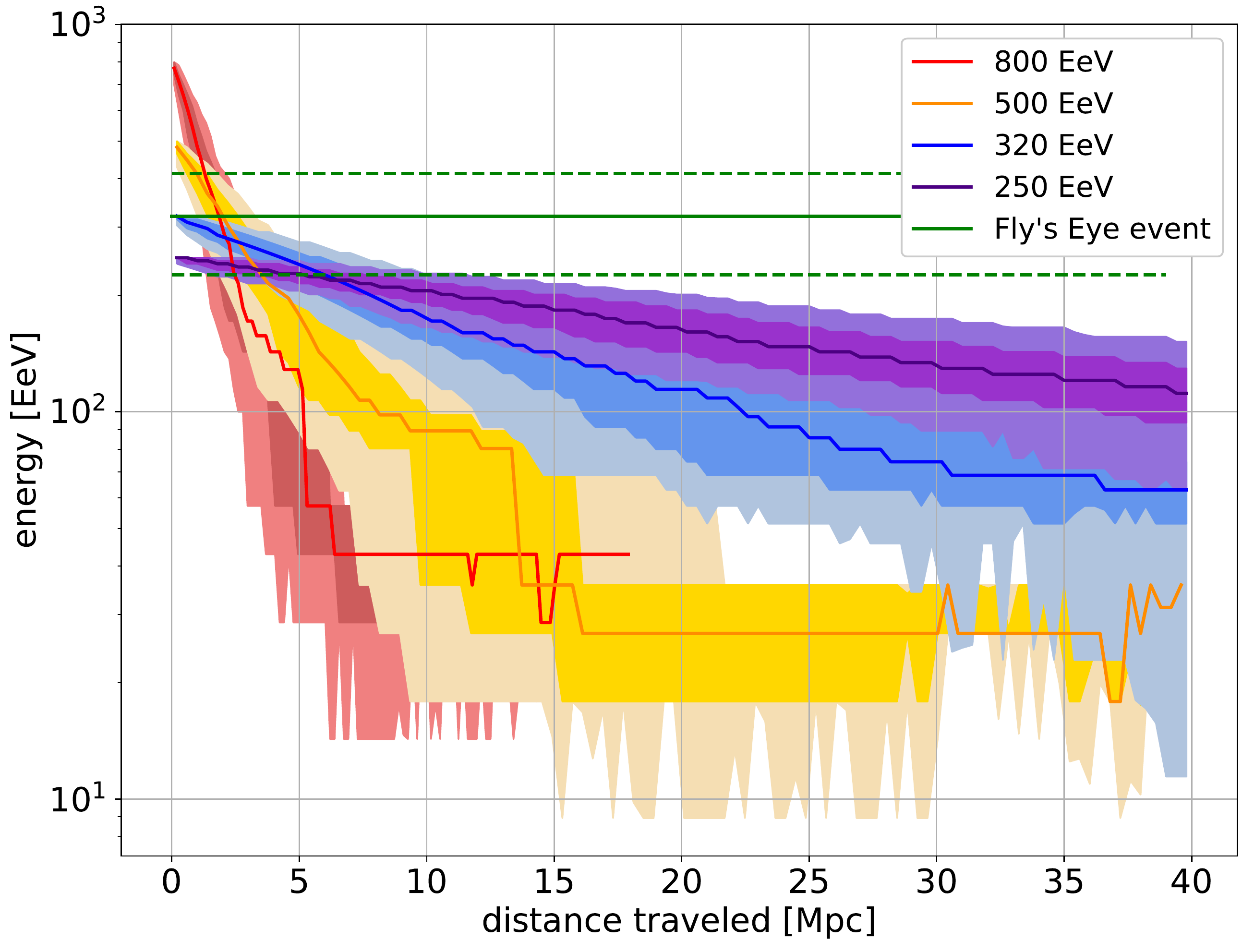} \\
    \includegraphics[width=0.8\textwidth]{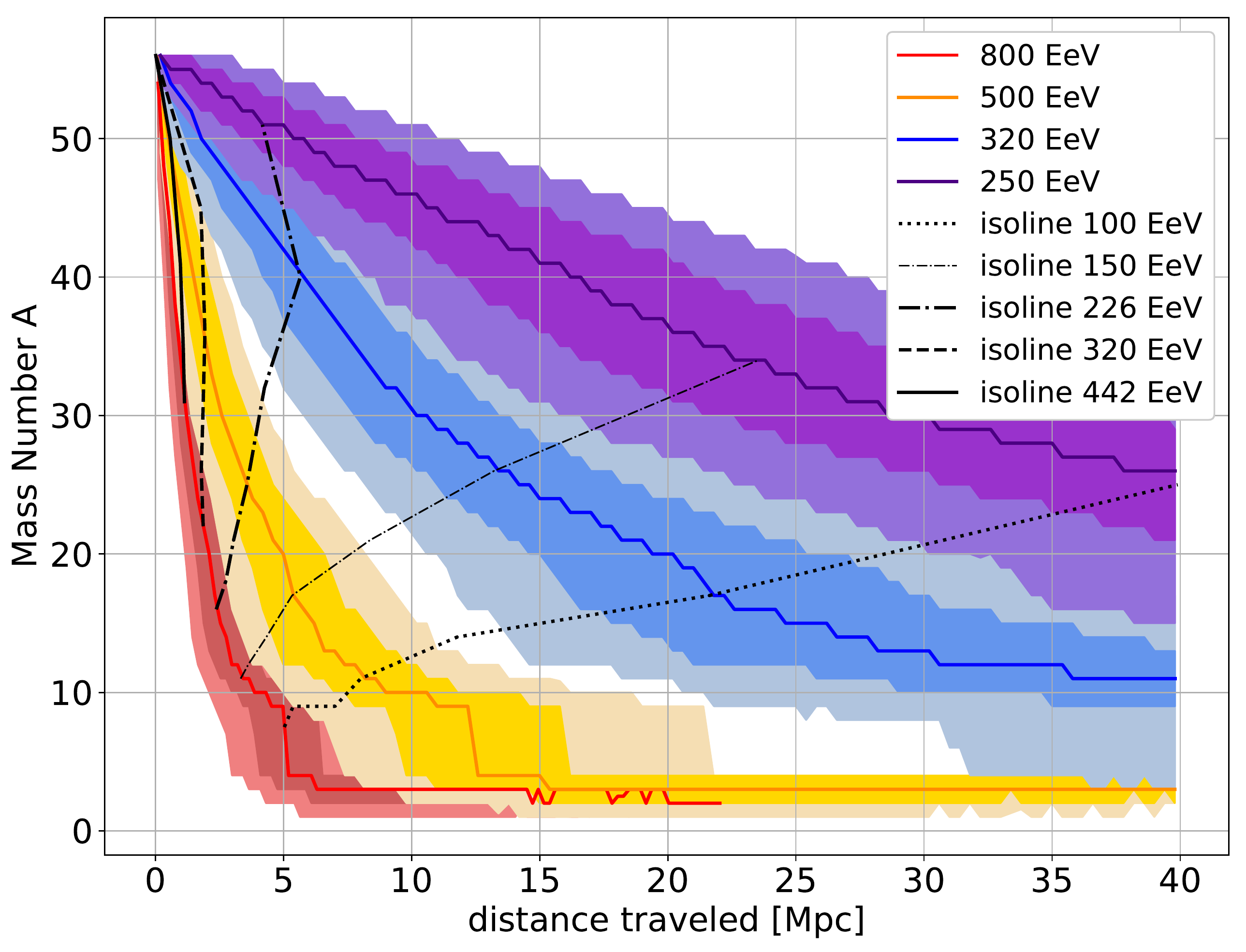}
    \caption{Evolution of the energy ({\bf a}) (top) and the mass number ({\bf b}) (bottom) of nuclei due to photo-disintegration for a primary iron nucleus injected at energies of 250 EeV (mauve), 320 EeV (blue), 500 EeV (yellow) and 800 EeV (red) versus traveled comoving distance ($D_{travel} = c\,t$, where $t$ is the comoving time). The solid lines are the median values, while the dark and light areas correspond to the 68\% and 95\% C.L. regions, respectively, for each injected energy. The isolines show where the particle reaches 100, 150, 226, 320 and 442 EeV, where 226 (420) EeV is the lower (upper) limit of the FE event.}
    \label{fig:Fe_AvsD}
\end{figure}

Using \CRPropa\ in 1D mode, we propagated $10^5$ iron nuclei for each injected energy at the source, discarding the cosmological effects. Photo-disintegration was computed off the photons of the cosmic microwave radiation and the infrared background (model from ref. \citep{kneiske_implications_2004}). Figure \ref{fig:Fe_AvsD} shows the evolution of (a) the energy and (b) the mass number of the leading particle versus the traveled distance, while losing nucleons through photo-disintegration. The Fe nuclei were injected at  250 (purple), 320 (blue), 500 (yellow) and 800 (red) EeV. The 68\% and 95\% confidence level are shown to highlight fluctuations and in Fig. \ref{fig:Fe_AvsD}b the isoenergy lines at 100, 226, 320 and 442 EeV are represented, where the second (fourth) value corresponds to the lower (upper) limit of the energy range of the FE event.

The first thing to be noted (see  Fig. \ref{fig:Fe_AvsD}a) is that, independently of the initial energy, the leading particle loses energy rapidly as a function of distance. Then, given the uncertainty in the energy for the FE event, as long as the injection energy is greater than the value measured on Earth, the source cannot be much farther than 5 Mpc, even taking into account fluctuations. Moreover, the median behavior favors a source at around 3 Mpc or less. 

Fig. \ref{fig:Fe_AvsD}b, on the other hand, shows that the change in mass composition has a strong dependence on the initial energy. An iron core emitted with an energy of 320 EeV, or less, can travel several tens of Mpc without the mass number of the nuclear fragments falling below 14. In contrast, above this energy, the mass of the leading particle degrades rapidly in less than 10 Mpc and, taking into account fluctuations, the mass of the arriving particle has a poor correlation with the injected energy. In addition, the isoenergy lines of the incoming particles indicate the existence of a pile-up, which implies a distance to the source of the FE event of no more than 5 Mpc and, most likely, about 3 Mpc.
As discussed in Fig. \ref{fig:Fe_AvsD}, propagation results in a pile-up of events inside the uncertainty energy range of the FE event around 320 EeV.

To establish this result in a more quantitative way and emphasize its relevance, we run a new set of simulations with CRPropa. This time we used a 3D mode with and without a magnetic field. We consider a magnetic field with an intensity of 1 nG and a coherence length of 1 Mpc, which are consistent with the extreme values that one can expect for the \EGMF. The Galactic magnetic field was discarded because at these energies it can barely deflect a particle, all the more true for the FE event, which is almost located towards the direction of the anti-galactic center. In order to reduce the computation time and increase the number of observed events, a source that emits isotropically was placed in the center of a sphere of radius equal to the distance from the source to the Earth. All the events reaching the sphere were recorded. The simulation was repeated for $2 \times 10^{5}$ Fe nuclei drawn from a power-law injection spectrum, between $10^{19}$ eV and $10^{22}$ eV. One simulation was performed for every set of parameters of distance, power-law index and magnetic field ($0$ or $1$ nG). 

In addition to the simulated secondary component, we generated a large set of events corresponding to the main diffuse component (the one observed so far by TA) so that the sum of both components produces a combined spectrum that fits the TA spectrum at lower energies (Eq. \ref{eq:J_TA}). The combined spectrum is scaled at $\log E = 20.25$, the energy
of the last point of TA measurements \citep{ivanov_energy_2012}.  In order to account for the uncertainty in the energy reconstruction, the events from the simulation are diluted with a Gaussian error $\sigma = \sqrt{0.21^2+0.10^2} = 0.23$  for TA (systematic and energy resolution errors \citep{ivanov_energy_2012}).

For each experiment, we drew randomly a set of events corresponding to the real number of events observed with $\log E \geq 19.3$. To take into account the energy reconstruction for HiRes and FE, the energy of the events are convolved with Gaussian errors corresponding to the difference between the statistical error from the given experiment and TA. This gives $\sigma  = \sqrt{\sigma_{FE}^2-\sigma_{TA}^2} = \sqrt{0.28^2-0.21^2} = 0.14$  for FE\ \cite{bird_detection_1995} and $\sigma = 0.12$ for HiRes \cite{belz_overview_2009}. One thousand independent realizations are produced in this way.

\begin{figure}[hbtp] \center
    \includegraphics[width=0.5\textwidth]{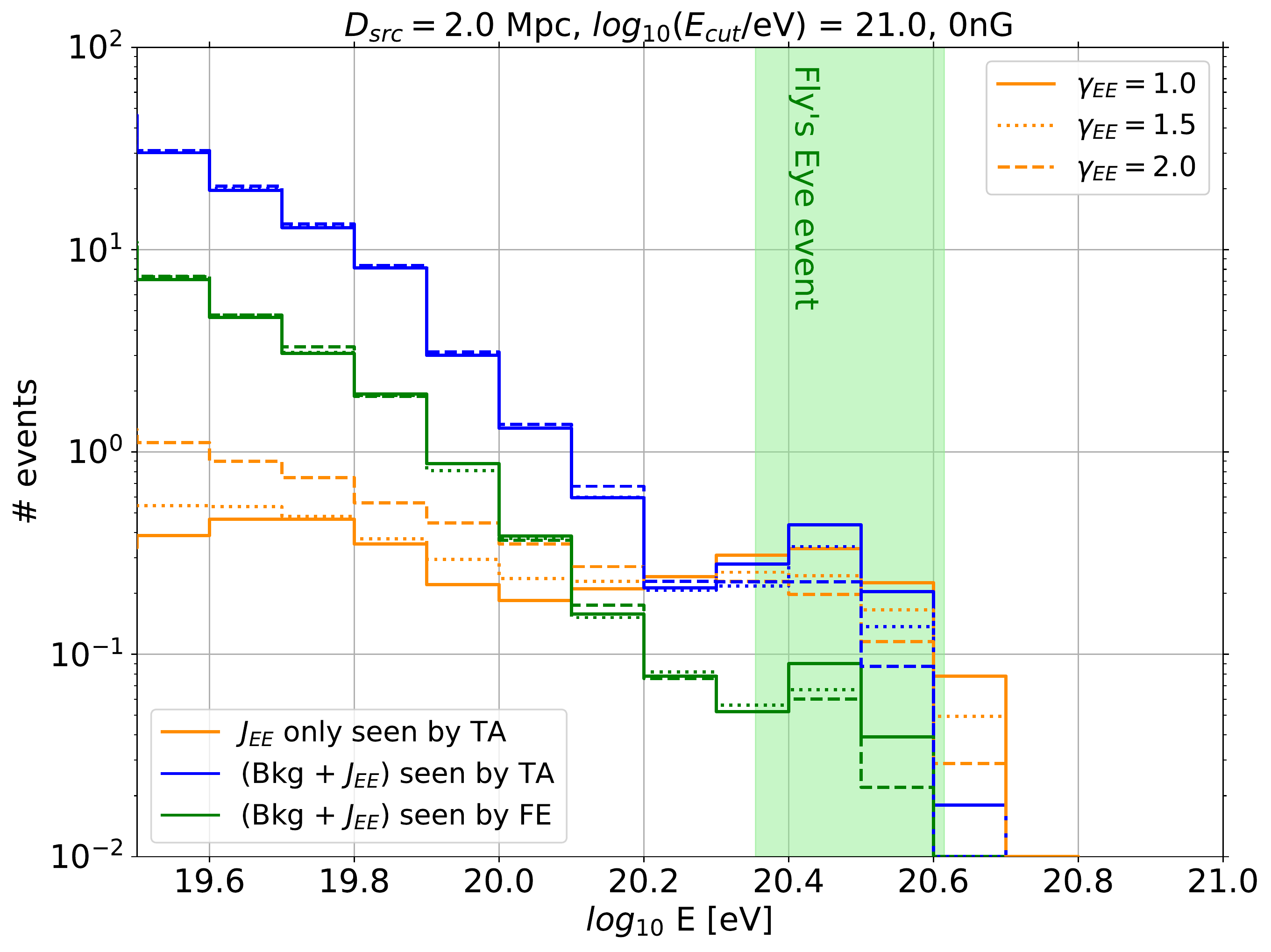} \\
    \includegraphics[width=0.5\textwidth]{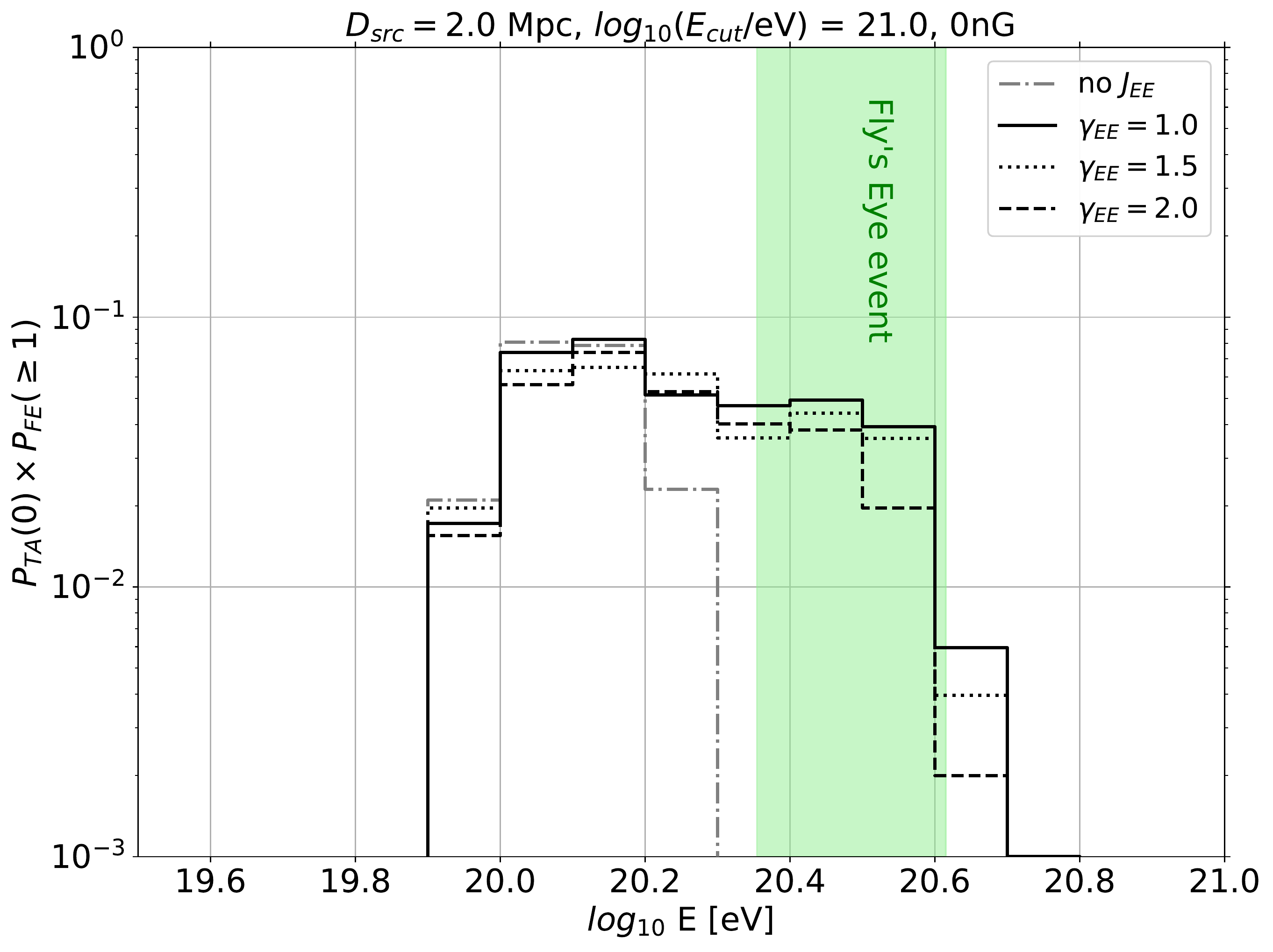} \\
    \includegraphics[width=0.5\textwidth]{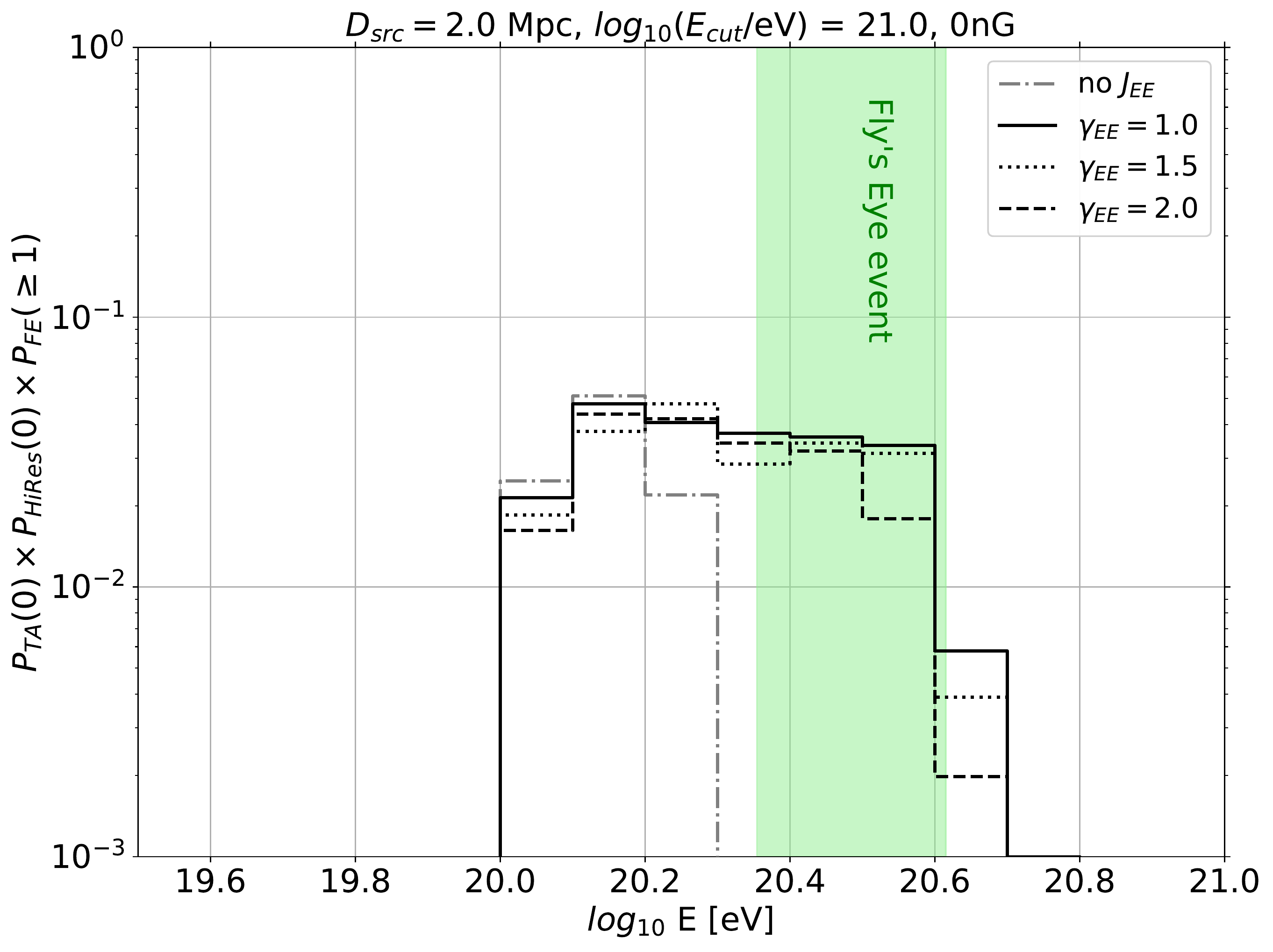}
    \caption{
    {\bf (a)} (top): Mean of the expected number of events for TA and FE (blue and green lines, respectively) under the assumption that besides the diffuse spectrum ($J_{D}$) there is an additional extreme energy component ($ J_{EE}$) originated from an iron source located at a distance of  2 Mpc. A power law spectrum with spectral index $\gamma_{EE}= 1.0, 1.5, 2.0$ is injected by the source, with a maximum energy at $E_{cut}=10^{21}$eV, and propagated through the diffuse photon background. This component at Earth is shown by the orange curves, normalized so that the combined spectrum ($J_{D} + J_{EE}$) fits the TA spectrum at lower energies.
{\bf (b)} (middle): Poissonian probability that FE detects at least one event ($P_{FE}(\geq1)$) and TA none ($P_{TA}(0)$) versus energy, considering that both experiments see the same spectrum for different spectral indexes of the heavy component: $\gamma_{EE}=1.0$ (solid lines), $1.5$ (dotted lines) and $2.0$ (broken lines).   {\bf (c)} (bottom): Idem as (b) but including also the probability of simultaneous non-observation by HiRes.} 
    \label{fig:Fe_spectrum}
\end{figure}

Fig. \ref{fig:Fe_spectrum} shows (a) the predicted number of events at the detector (averaged over the 1000 realizations), as well as the probability as a function of energy of observation by FE and simultaneous non-observation by either (b) TA alone or (c) TA and HiRes combined (see the figure caption for details). In accordance with the previous discussions, we assume that the observed spectrum includes a heavy (Fe) component of extreme energy originated in a bursting source located at a distance of 2 Mpc, as suggested in Figure \ref{fig:Fe_AvsD}. This Fe component escapes the source as a power low with varying spectral indexes $\gamma_{EE}=$ 1, 1.5 and 2. We performed 1000 independent realizations for each detector. 

From the results of these simulations, we can compute the Poissonian probabilities $P_{TA}(0)\times P_{FE}(\geq1)$ and $P_{TA}(0) \times P_{HiRes}(0) \times P_{FE}(\geq1)$ as functions of energy, by counting for each energy bin the number of times, among the 1000 independent realizations of the observed spectra, that 0 events were observed by TA and HiRes and at least 1 event was observed by FE. This gives us directly the probability of reproducing the observations of these three experiments. As shown in Figs. \ref{fig:Fe_spectrum}b 
and \ref{fig:Fe_spectrum}c, for each value of the parameter $\gamma_{EE}$, a peak appears in the region corresponding to the energy uncertainty interval of the FE event, which renders the observations plausible.  If only TA and FE are considered, the probabilities of TA non-observation and FE observation is $\sim 5.5$\% for a very hard spectrum, $\gamma_{EE}=1$, $\sim 4.5$\% for $\gamma_{EE}=1.5$ and $\sim 4$\% for $\gamma_{EE}=2$. While the probabilities decrease when HiRes is added, their respective values are still non-negligible: $\sim 4$\%, 4\% and 3.5\%.

It must be noted that the injection of hard spectra at the source is not unreasonable. In fact, a combined spectrum and composition fit with 5 nuclear species performed by the Auger Collaboration \cite{PAO_HardSpceFit}, favors low spectral indexes ($\gamma \lesssim 1$). This may seem in contradiction with the expectations from the most plausible acceleration mechanisms, e.g., 1st order Fermi acceleration,  which predicts  power-laws with $\gamma_{acc} \gtrsim 2$. But there is no contradiction if a distinction is made between the acceleration spectrum and the spectrum of the particles that are actually able to escape from the source into the intergalactic medium. In fact, photo-disintegration in the surroundings of UHECR sources has been shown to harden the spectrum to values of the spectral index as small as $\gamma_{esc} \sim 1$ \cite{Unger_SourceEscape_HardSpec}, while acceleration models involving Gamma-Ray Bursts (GRBs), unipolar inductors, magnetic reconnection, and tidal disruption events have been suggested as producing hard spectra with $0 \leq \gamma_{acc} \leq 1.5$  
\cite{Unger_SourceEscape_HardSpec, Globus_GRB_HardSpec, Baerwald_GRB_HardSpec, Biehl_GRB_HardSpec, Zhang_GRB_HardSpec, Blasi_YNSW_HardSpec, Fang_NBPulsars_HardSpec, Kowal_Reconnection_HardSpec, Drury_1stOFMReconnection_HardSpec, Alves_TidalDisruption_HardSpec, Bihel_TidalDisruption_HardSpec, Zhang_TidalDisruption_HardSpec, Guepin_TidalDisruption_HardSpec}. 

Hence, the previous result strongly argues for a northern sky source of heavy nuclei, injecting a hard spectrum (either due to acceleration or escape mechanisms) located at $\sim 2 - 3$ Mpc, as a plausible explanation for the FE event combined with a null result of TA at the same energy. Such a potent source in the immediate neighborhood of the Local Group must be transient, and far enough in the past as to be unobservable in high energy photons at present. Therefore, stellar mass range candidates, e.g. GRB, are favored. 

In this scenario, the source of the FE event could be just the same type of source responsible for the general UHECR flux seen in both hemispheres, but this particular one just happened to burst nearby and in a position of the sky where only a northern instrument could see it. Therefore, the existence of such source would not challenge the current interpretation of observations either by TA or PAO: other sources of the same population would be just far away and reproduce collectively a spectrum with the suppression and other spectral features consistent with observations by both modern experiments. 

A preliminary search for potential sources in the vicinity of the event did not produce any galaxy that could host such event. The closest galaxy at less than 15$^\circ$ from the FE event location is PGC 019156, which is located at 13 Mpc, too far according to our analysis. Galaxies in the Local Group at less than 5 Mpc are at least 25$^\circ$ from the FE event. The absence of obvious candidate hosting the source may be an indication of a deeper physical process or a poor knowledge of the magnetic field inside the Local Group. A detailed study about this latter possibility is in progress.

\section{FE event as a proton}
\label{sec:proton}

Protons don't seem to be favored at present as a major component of the UHECR flux at the highest energies. Nevertheless, they cannot be ruled out as part of the flux coming from some specific source mechanism. In any case, they indeed must be present at least as fragments of photo-disintegration. Therefore, it is worth exploring protons as a candidate for the 
FE event. 

\begin{figure}[hbtp] \center
    \includegraphics[width=0.8\textwidth]{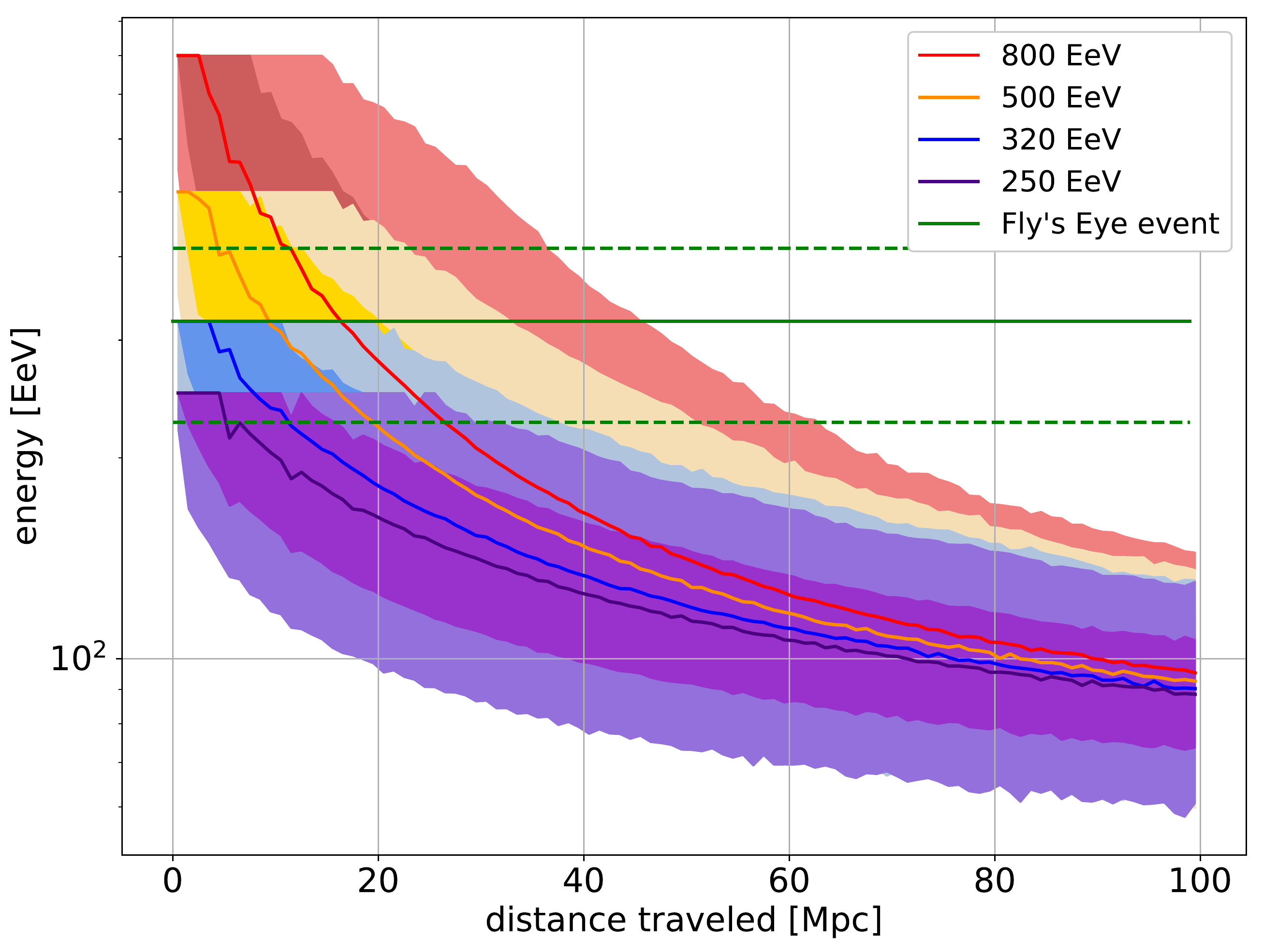}
    \caption{Energy evolution due to photo-pion production of protons with initial energies of 100 EeV (mauve), 320 EeV (blue), 500 EeV (yellow) and 800 EeV (red) versus the traveled cosmological distance ($D_{travel} = c\,t$, where $t$ is the cosmological time). The solid lines are the median values, while the dark and light areas correspond to the 68\% and 95\% C.L. regions, respectively, for each injected energy.}
    \label{fig:H_EvsD}
\end{figure}

The same analysis made in Sec. \ref{sec:iron} is repeated here for a source injecting protons. Fig. \ref{fig:H_EvsD} shows the evolution of the energy of a proton due to photo-pion production versus traveled distance. The median value and the corresponding 68 and 95 \% C.L. are shown for four different initial energies: 250, 320, 500 and 800 EeV. The method is the same as the one used for Fig. \ref{fig:Fe_AvsD}a. The energy limits of the FE event are also plotted.
First, one can note that, due to statistical fluctuations, to the energies of the FE event, 16\% of the protons can travel between 10 and 20 Mpc without interaction, and 2.5\% an even longer distance. Particles injected at higher energies still have a considerable probability to reach Earth inside the FE event energy uncertainty band, from distances of several tens of Mpc. Increasing the energy with which the proton is injected only steadily increases the distance at which the FE event energy is reached (fluctuations included). No particular distance scale emerges as favored, with the possible exception of 20 Mpc, which englobes 95\% probability of particles arriving without interaction. 

\begin{figure}[hbtp] \center
    \includegraphics[width=0.5\textwidth]{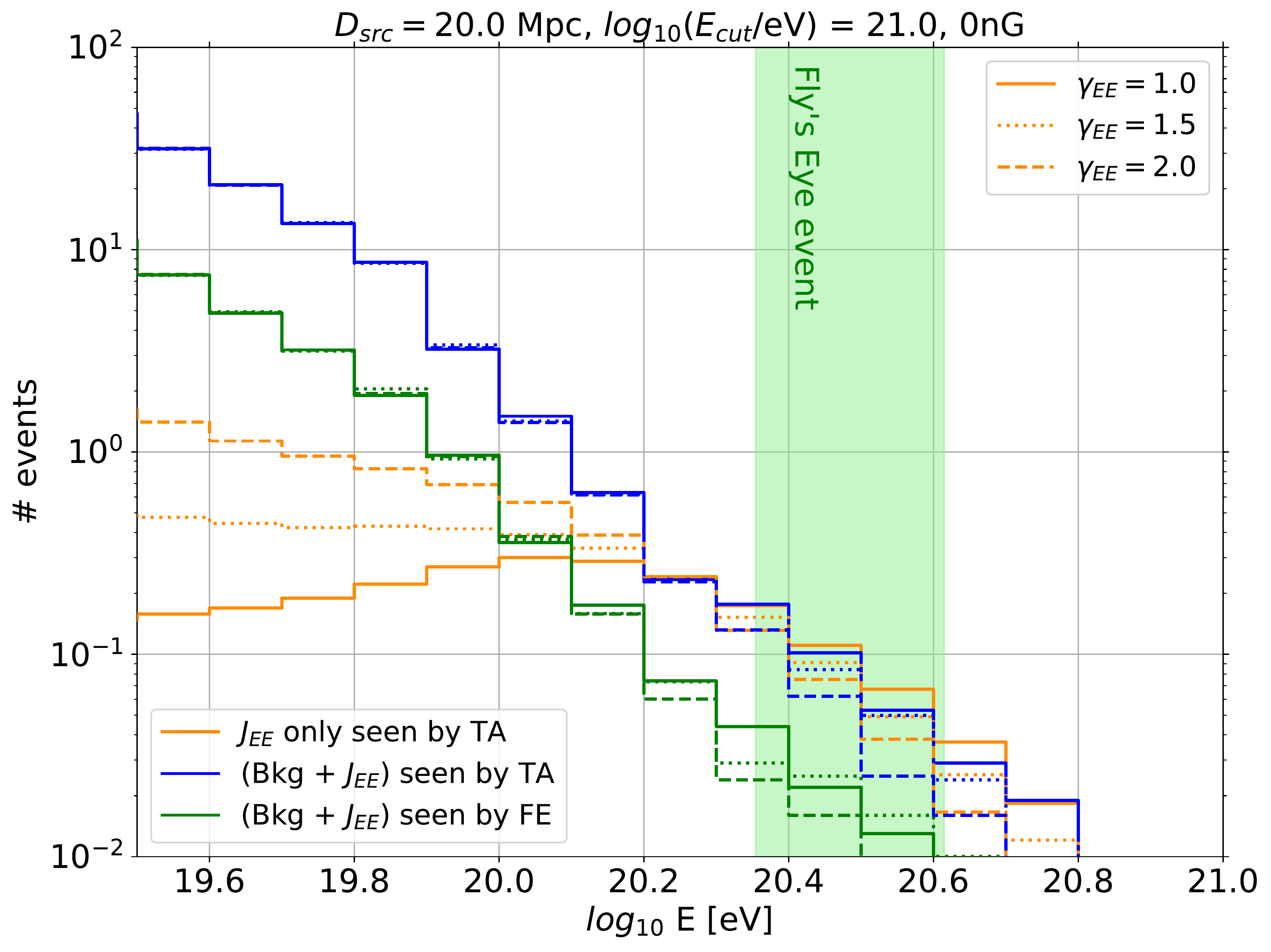} \\
    \includegraphics[width=0.5\textwidth]{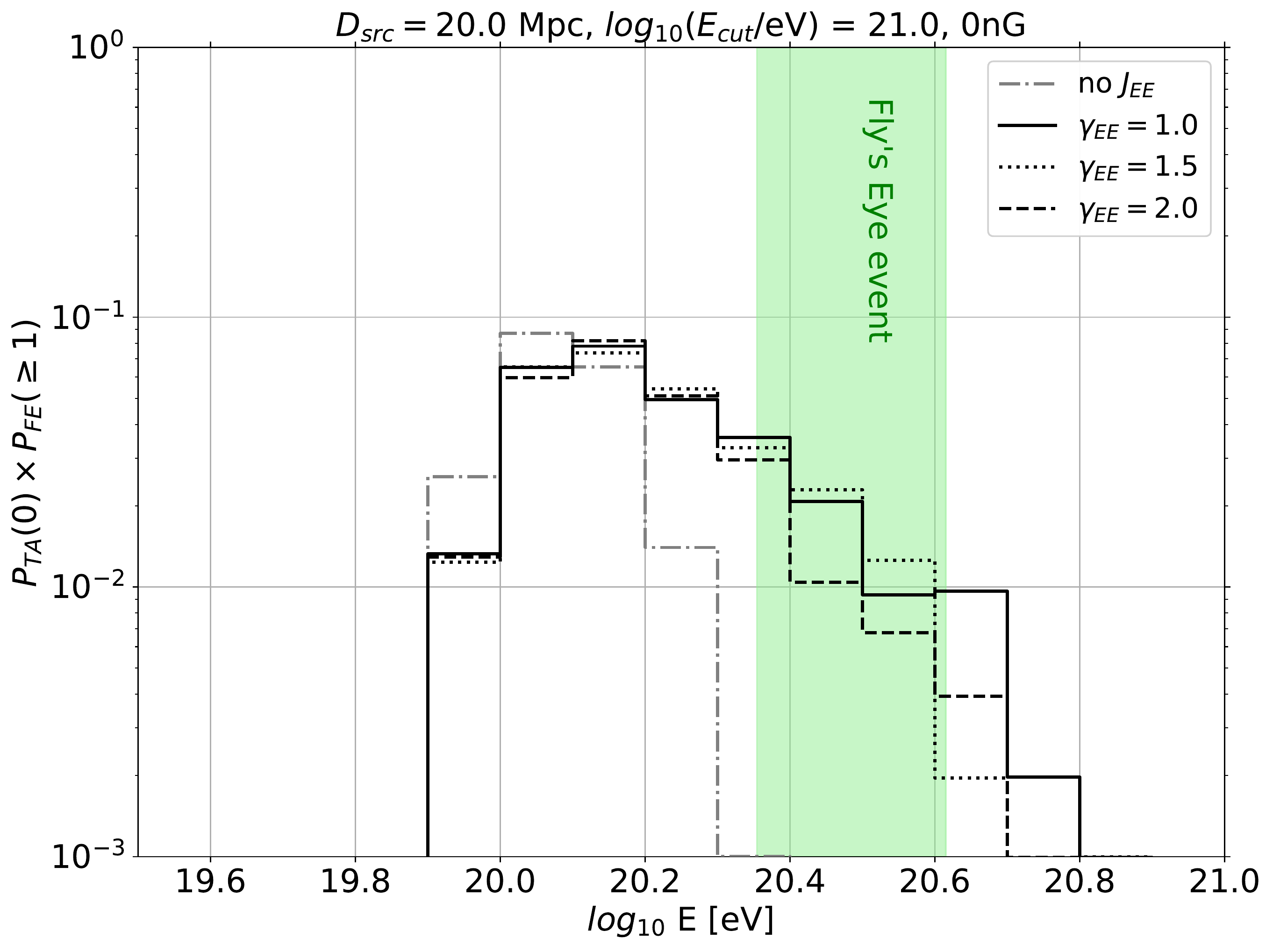}
    \includegraphics[width=0.5\textwidth]{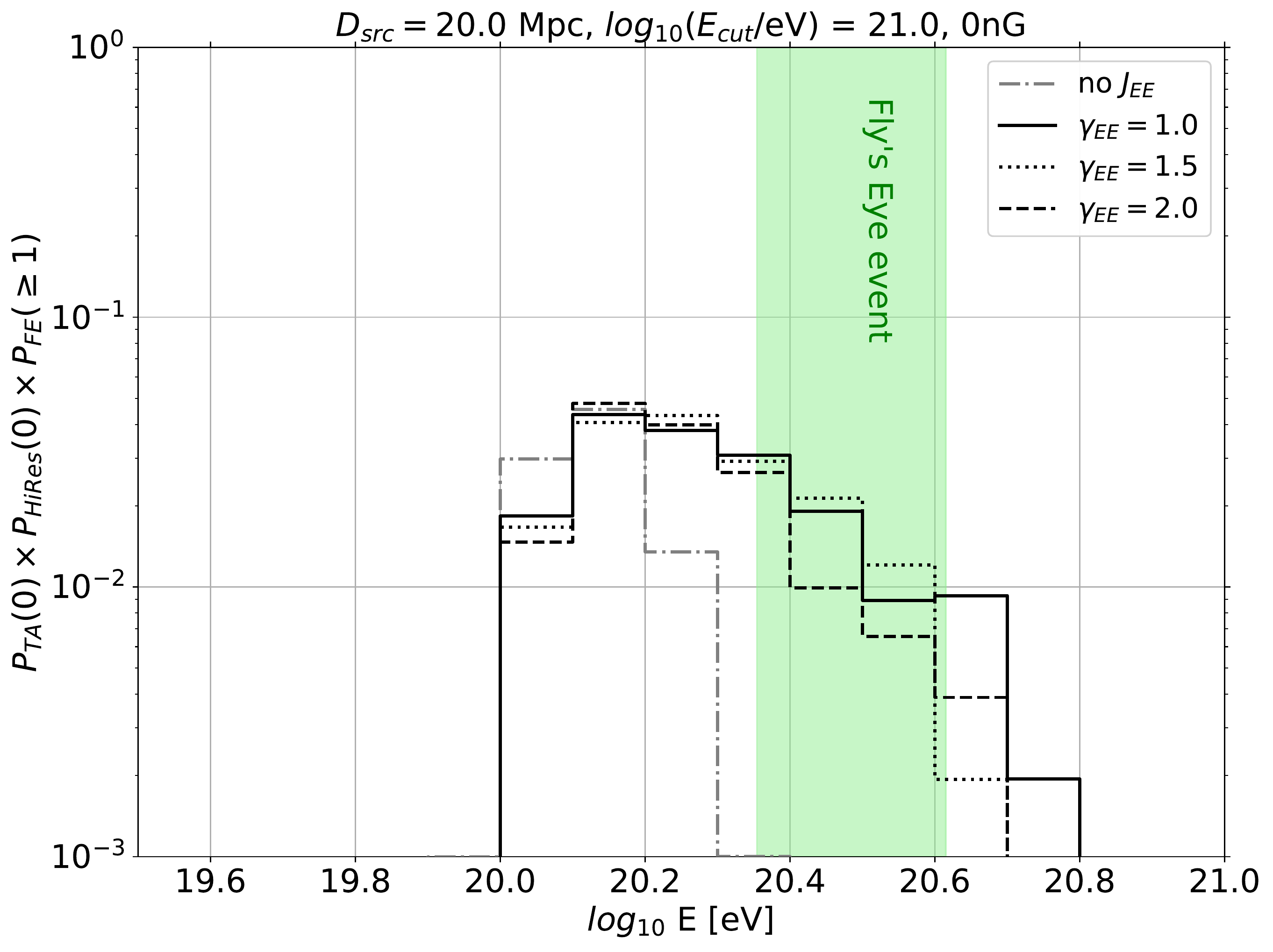}
    \caption{
     {\bf (a)} (top): Mean of the expected number of events for TA and FE (blue and green lines, respectively) under the assumption that besides the diffuse spectrum ($J_{D}$) there is an additional extreme energy component ($ J_{EE}$) originated from a proton source located at a distance of 20 Mpc. A power law spectrum with spectral index $\gamma_{EE}= 1.0, 1.5, 2.0$ is injected by the source, with a maximum energy at $E_{cut}=10^{21}$eV, and propagated through the diffuse photon background. This component at Earth is shown by the orange curves, normalized so that the combined spectrum ($J_{D} + J_{EE}$) fits the TA spectrum at lower energies.
{\bf (b)} (middle): Poissonian probability that FE detects at least one event ($P_{FE}(\geq1)$) and TA none ($P_{TA}(0)$) versus energy, considering that both experiments see the same spectrum for different spectral indexes of the heavy component: $\gamma_{EE}=1.0$ (solid lines), $1.5$ (dotted lines) and $2.0$ (broken lines).   {\bf (c)} (bottom): Idem as (b) but including also the probability of simultaneous non-observation by HiRes.}     
    \label{fig:H_spectrum}
\end{figure}

Fig. \ref{fig:H_spectrum} is the proton equivalent  to Fig.  \ref{fig:Fe_spectrum} but for a source at 20 Mpc. One can easily recognize that, unlike the Fe injection, no pattern appears. In particular, from Fig. \ref{fig:H_spectrum}b it can be seen that the only maximum of $P_{TA}(0)\times P_{FE}(\geq1)$ occurs at $E=10^{20.1}$ eV, independently of the hardness of the injected spectrum. This a remarkable difference with respect to the Fe injection, which has a peak around $\sim 320$ EeV, consistent with the FE event.

Thus, contrary to the heavy nuclei scenario, no obvious specific pattern, natural distance or energy scale arises which can favor the FE event at 320 EeV and the simultaneous non-observation of events between 100 and 320 EeV by TA. Furthermore, the source should be relatively far away, tens of Mpc, making it harder to justify the absence of another such source in the field of view of PAO. Additionally, if it were indeed a separate kind of source emitting high energy protons, those protons should also be observable at both hemispheres, as a distinct light component, in the general composition measurement at lower energies. Thus, though not conclusively, protons do not seem to be favored as a candidate for the FE event. 

\section{Discussion}
\label{sec:conclusion}

Extrapolating the current measurements, we show that the TA spectrum by itself cannot provide a reasonable explanation for the observation of the FE event. A secondary component with a hard spectrum is required. But this alone, without considering the nature of the primary particle involved, is not enough to explain why FE detected an event with extremely high energy ($\sim 320$ EeV) and, at the same time, why the TA detector, with an exposure 10 times greater than the one of FE, has not observed anything at those energies.

With the aim of taking into account these important characteristics of the observations, in this work we study the propagation of iron nuclei and protons. Based on the calculations, we show that iron nuclei injected with energies greater than 320 EeV tend to reach this energy after traveling distances between 2 and 3 Mpc, while the mass of the main fragment vary from N to Fe. As a consequence, a source located at such distance and that emits a power law spectrum with a limit of $ 10 ^ {21} $ eV generates a spectrum on Earth with an excess of events in the region around the energy of the FE event.

Since the incoming direction of the event is within the field of vision of both experiments, besides the mere existence of the 320 EeV event, we faced the problem of understanding why more advanced experiments like TA have not observed a similar event. Contrary to normal intuition, we demonstrated that the Fe pile-up results in a combined probability of both results, $P_{TA}(0)\times P_{FE}(\geq1)$ which can be as high as 5\% for a hard spectrum with $\gamma_{EE} \sim 1$.  In this sense, it is important to remark that a like hard spectrum ($\gamma_{EE}=1$) and a low cut-off  energy ($10^{21}$ eV) are compatible with recent experimental results \citep{the_pierre_auger_collaboration_combined_2017,batista_cosmogenic_2019}. The small power law index does not necesarily have to be a consequence of the acceleration mechanism, but rather the result of the interaction of the nuclei with the environment of the source during the escape process.

A prominent Galactic source able to accelerate particle beyond $10^{20}$ eV is not apparent inside the error box of the FE event, neither in $\gamma$-rays nor $X$-rays. Therefore, it is more likely to be a transient source whose activity ended in the past, a time long enough that only the component of charged particles is arriving today. At such high energies the Galactic magnetic field would only add a few degrees to an already large FE error box, inside of which there are no conspicuous X-ray sources which could be capable to accelerate particles up to 300 EeV \citep[c.f., the Chandra Catalog][]{evans_thechandra_2010} and, in this way, significantly change our previous statement. The bursting source should probably be related to an object that is the product of stellar evolution. Due to its spatial and angular location, the source would only be visible in the Northern Hemisphere, outside the field of view of the PAO, explaining the fact that there is no counterpart of the FE event in this detector. 

In the event that the primary was a proton, we demonstrate that the highest probability of detecting a particle with FE and not with TA is at $\sim $ 100 EeV instead of the $ 320 $ EeV observed, which makes it difficult to justify the existence of the FE event together with a null results from HiRes and TA. Moreover, there is no reason why a source of protons should be located only in the Northern Hemisphere and at a short distance, which conflicts with the lack of detection by the PAO. In addition, the associated neutrino and gamma-rays productions would also be important and can exceed the limits already set on the neutrino and gamma-ray background \citep {batista_cosmogenic_2019,van_vliet_determining_2019}. As a consequence, even though the possibility of a secondary pure proton component with a hard spectrum extending at higher energy cannot be completely ruled out, it looks much more complicated to implement.

Finally, let's briefly comment on the viability of a photon as a primary particle. In principle, this cannot fully disregarded. We can envision two possible scenarios: a conservative bottom-up and a more exotic top-down. For a bottom-up scenario, an extragalactic source seems unlikely due to the opacity of the Universe to high energy photons. If the source were Galactic, then it would be very difficult to justify that the only messengers produced are UHE photons. In fact, an energetic source should also be a very luminous source of photons at lower energies and be easily identifiable by several astronomical observatories within the error box of the FE event. To our knowledge this is not the case. Regarding to the top-down mechanism, in theory, superheavy dark matter is not excluded. Nevertheless, the FE event error box is clearly located toward the outskirts of the Galaxy, where the dark matter density is much smaller that in the center. Since the center of the Galaxy is in the full view of the PAO, with a much larger exposure and no clear signature of UHE photons, it does not seem that this scenario is favored.

\acknowledgments

This work was supported by DGAPA-UNAM Grant PAPIIT IN103207 and by CONACYT Grants CB 239660 and 
CB 240666. The authors acknowledge Mtro. Juan Luciano D\'iaz Gonz\'alez for his technical assistance with the 
computational cluster  infrastructure. One of the authors (TF) would like to thank  H\'el\`ene Courtois for her help 
in trying to locate a potential source of the FE event. TF also acknowledges support from DGAPA-UNAM.

\bibliographystyle{JHEP}
\bibliography{Legacy_from_Flys_Eye_JCAP}

\providecommand{\href}[2]{#2}\begingroup\raggedright\begin{thebibliography}{10}

\bibitem{bird_detection_1995}
D.~J. Bird et~al., \emph{Detection of a cosmic ray with measured energy well
  beyond the expected spectral cutoff due to cosmic microwave radiation},
  \href{https://doi.org/10.1086/175344}{\emph{ApJ} {\bfseries 441} (1995) 144}.

\bibitem{greisen_end_1966}
K.~Greisen, \emph{End to the cosmic-ray spectrum?},
  \href{https://doi.org/10.1103/PhysRevLett.16.748}{\emph{Phys. Rev.~Lett.}
  {\bfseries 16} (1966) 748}.

\bibitem{zatsepin_upper_1966}
G.~T. Zatsepin and V.~A. Kuz'min, \emph{Upper limit of the spectrum of cosmic
  rays}, {\emph{Soviet. Phys. - JETP Letters} {\bfseries 4} (1966) 78}.

\bibitem{allard_extragalactic_2012}
D.~Allard, \emph{Extragalactic propagation of ultrahigh energy cosmic-rays},
  \href{https://doi.org/10.1016/j.astropartphys.2011.10.011}{\emph{Astroparticle
  Physics} {\bfseries 39-40} (2012) 33}.

\bibitem{risse_primary_2004}
M.~Risse, \emph{Primary particle type of the most energetic {Fly's Eye} air
  shower},
  \href{https://doi.org/10.1016/j.astropartphys.2004.04.003}{\emph{Astroparticle
  Physics} {\bfseries 21} (2004) 479}.

\bibitem{risse_primary_2006}
M.~Risse et~al., \emph{On the primary particle type of the most energetic
  {Fly's Eye} event},
  \href{https://doi.org/10.1016/j.nuclphysbps.2005.07.017}{\emph{Nuclear Phys.
  B - Proc. Sup.} {\bfseries 151} (2006) 96}.

\bibitem{durrer_cosmological_2013}
R.~Durrer and A.~Neronov, \emph{Cosmological magnetic fields: their generation,
  evolution and observation},
  \href{https://doi.org/10.1007/s00159-013-0062-7}{\emph{A\&ARv} {\bfseries 21}
  (2013) 62}.

\bibitem{sigl_origin_1994}
G.~Sigl, D.~N. Schramm and P.~Bhattacharjee, \emph{On the origin of highest
  energy cosmic rays},
  \href{https://doi.org/10.1016/0927-6505(94)90029-9}{\emph{Astroparticle
  Physics} {\bfseries 2} (1994) 401}.

\bibitem{elbert_search_1995}
J.~W. Elbert and P.~Sommers, \emph{In search of a source for the 320 {EeV}
  {Fly's Eye} cosmic ray}, \href{https://doi.org/10.1086/175345}{\emph{ApJ}
  {\bfseries 441} (1995) 151}.

\bibitem{rachen_possible_1995}
J.~P. Rachen, \emph{Possible extragalactic sources of the highest energy cosmic
  rays}, \href{https://doi.org/10.1111/j.1749-6632.1995.tb17587.x}{\emph{Annals
  of the New York Academy of Sciences} {\bfseries 759} (1995) 468}.

\bibitem{milgrom_possible_1995}
M.~Milgrom and V.~Usov, \emph{Possible association of ultra--high-energy
  cosmic-ray events with strong gamma-ray bursts},
  \href{https://doi.org/10.1086/309633}{\emph{The Astrophysical Journal}
  {\bfseries 449} (1995) L37}.

\bibitem{gnatyk_search_2016}
R.~Gnatyk, Y.~Kudrya and V.~Zhdanov, \emph{Search for the astrophysical sources
  of the {Fly's Eye} event with the highest to date cosmic ray energy e=$3.2
  \cdot 10^{20}$ {eV}},
  \href{https://doi.org/10.17721/2227-1481.6.41-44}{\emph{Advances in Astronomy
  and Space Physics} {\bfseries 6} (2016) 41}.

\bibitem{bird_cosmic-ray_1994}
D.~J. Bird, S.~C. Corbató, H.~Y. Dai, B.~R. Dawson, J.~W. Elbert, T.~K. Gaisser
  et~al., \emph{The cosmic-ray energy spectrum observed by the fly's eye},
  \href{https://doi.org/1994ApJ...424..491B}{\emph{The Astrophysical Journal}
  (1994) }.

\bibitem{bird_results_1995}
D.~J. Bird et~al., \emph{Results from the {Fly's Eye} experiment},  in
  \emph{{AIP} Conference Proceedings}, pp.~839--854, {AIP}, 1995,
  \href{https://doi.org/10.1063/1.48457}{DOI}.

\bibitem{bergman_cosmic_2007}
D.~R. Bergman and J.~W. Belz, \emph{Cosmic rays: the second knee and beyond},
  \href{https://doi.org/10.1088/0954-3899/34/10/R01}{\emph{J. Phys. G: Nucl.
  Part. Phys.} {\bfseries 34} (2007) R359}.

\bibitem{bergman_observation_2007}
D.~R. Bergman, \emph{Observation of the {GZK} cutoff using the {HiRes}
  detector},
  \href{https://doi.org/10.1016/j.nuclphysbps.2006.11.004}{\emph{Nuclear
  Physics B - Proceedings Supplements} {\bfseries 165} (2007) 19}.

\bibitem{abbasi_first_2008}
M.~Abbasi, R. U.~Abe, T.~Abu-Zayyad et~al., \emph{First observation of the
  {Greisen-Zatsepin-Kuzmin} suppression},
  \href{https://doi.org/10.1103/PhysRevLett.100.101101}{\emph{Phys. Rev.~Lett.}
  {\bfseries 100} (2008) }.

\bibitem{belz_overview_2009}
J.~Belz, \emph{Overview of recent {HiRes} results},
  \href{https://doi.org/10.1016/j.nuclphysbps.2009.03.061}{\emph{Nuclear
  Physics B - Proceedings Supplements} {\bfseries 190} (2009) 5}.

\bibitem{ivanov_energy_2012}
D.~Ivanov, \emph{Energy spectrum measured by the telescope array surface
  detector},  2012.
\newblock PhD thesis.

\bibitem{the_telescope_array_collaboration_pierre_2018}
T.~T. Collaboration and T.~P. Collaboration, \emph{{Pierre Auger} observatory
  and {Telescope Array}: Joint contributions to the 35th international cosmic
  ray conference ({ICRC} 2017)}, {\emph{{arXiv}:1801.01018 [astro-ph]} (2018) }
  [\href{https://arxiv.org/abs/1801.01018}{{\ttfamily 1801.01018}}].

\bibitem{PAOandTA_2018}
J.~Biteau et~al., \emph{Covering the celestial sphere at ultra-high energies:
  full-sky cosmic-ray maps beyond the ankle and the flux sup- pression},
  {\emph{Proceedings of Ultra High Energy Cosmic Rays 2018} (2018) 8}.

\bibitem{sokolsky_observation_2009}
P.~Sokolsky, \emph{Observation of the {GZK} cutoff by the {HiRes} experiment},
  \href{https://doi.org/10.1016/j.nuclphysbps.2009.09.010}{\emph{Nuclear
  Physics B - Proceedings Supplements} {\bfseries 196} (2009) 67}.

\bibitem{verzi_measurement_2017}
V.~Verzi, D.~Ivanov and Y.~Tsunesada, \emph{Measurement of energy spectrum of
  ultra-high energy cosmic rays}, {\emph{{ArXiv} e-prints} (2017)
  arXiv:1705.09111}.

\bibitem{abbasi_indications_2010}
M.~Abbasi, R. U.~Abe, T.~Abu-Zayyad et~al., \emph{Indications of
  proton-dominated cosmic ray composition above 1.6 {EeV}},
  \href{https://doi.org/10.1103/PhysRevLett.104.161101}{\emph{Phys. Rev.~Lett.}
  {\bfseries 104} (2010) } [\href{https://arxiv.org/abs/0910.4184}{{\ttfamily
  0910.4184}}].

\bibitem{PAO_TA_comp_2016}
{Pierre Auger} and T.~Collaborations, \emph{Report of the working group on the
  composition of ultrahigh energy cosmic ray}, {\emph{JPS Conf. Proc. 9, 010016
  (2016)} (2016) doi:10.7566/JPSCP.9.010016 [arXiv:1503.07540]}.

\bibitem{Hanlon_comp_2018}
H.~Hanlon et~al., \emph{Report of the working group on the mass composition of
  ultrahigh energy cosmic rays}, {\emph{JPS Conf. Proc. 19, 011013 (2018)}
  (2018) doi:10.7566/JPSCP.19.011013}.

\bibitem{batista_crpropa_2016}
R.~A. Batista, A.~Dundovic, M.~Erdmann et~al., \emph{{CRPropa} 3 - a public
  astrophysical simulation framework for propagating extraterrestrial
  ultra-high energy particles},
  \href{https://doi.org/10.1088/1475-7516/2016/05/038}{\emph{J.
  Cosmol.~Astropart.~Phys.} {\bfseries 2016} (2016) 038}.

\bibitem{kneiske_implications_2004}
T.~M. Kneiske, T.~Bretz, K.~Mannheim and D.~H. Hartmann, \emph{Implications of
  cosmological gamma-ray absorption: {II}. modification of gamma-ray spectra},
  \href{https://doi.org/10.1051/0004-6361:20031542}{\emph{A\&A} {\bfseries 413}
  (2004) 807}.

\bibitem{PAO_HardSpceFit}
P.~Collaboration, \emph{Combined fit of spectrum and composition data as
  measured by the {Pierre Auger} observatory}, {\emph{Journal of Cosmology and
  Astroparticle Physics 4 (Apr., 2017) 038} (2017) arXiv:1612.07155}.

\bibitem{Unger_SourceEscape_HardSpec}
M.~Unger, G.~R. Farrar and L.~A. Anchordoqui, \emph{Origin of the ankle in the
  ultrahigh energy cosmic ray spectrum and of the extragalactic protons below
  it}, {\emph{Physical Review D 92 (Dec., 2015) 123001} (2015)
  arXiv:1505.02153}.

\bibitem{Globus_GRB_HardSpec}
N.~Globus, D.~Allard, R.~Mochkovitch and E.~Parizot, \emph{{UHECR} acceleration
  at {GRB} internal shocks}, {\emph{Monthly Notices of the Royal Astronomical
  Society 451 (July, 2015) 751?790} (2015) arXiv:1409.1271}.

\bibitem{Baerwald_GRB_HardSpec}
P.~Baerwald, M.~Bustamante and W.~Winter, \emph{{UHECR} escape mechanisms for
  protons and neutrons from {GRB}s, and the cosmic ray-neutrino connection},
  {\emph{Astrophys. J. 768 (2013) 186} (2013) arXiv:1301.6163}.

\bibitem{Biehl_GRB_HardSpec}
D.~Biehl, D.~Boncioli, A.~Fedynitch and W.~Winter, \emph{Cosmic-ray and
  neutrino emission from gamma-ray bursts with a nuclear cascade},
  {\emph{Astron. Astrophys. 611 (2018) A101} (2017) arXiv:1705.08909}.

\bibitem{Zhang_GRB_HardSpec}
B.~T. Zhang, K.~Murase, S.~S. Kimura, S.~Horiuchi and P.~M\'esz\'aros,
  \emph{Low-luminosity gamma-ray bursts as the sources of ultrahigh-energy
  cosmic ray nuclei}, .

\bibitem{Blasi_YNSW_HardSpec}
P.~Blasi, R.~I. Epstein and A.~V. Olinto, \emph{Ultra-high-energy cosmic rays
  from young neutron star winds}, {\emph{The Astrophysical Journal Letters 533
  (Apr., 2000) L123?L126} (2000) arXiv:9912240}.

\bibitem{Fang_NBPulsars_HardSpec}
K.~Fang, K.~Kotera and A.~V. Olinto, \emph{Newly born pulsars as sources of
  ultrahigh energy cosmic rays}, {\emph{The Astrophysical Journal 750 (May,
  2012) 118} (2012) arXiv:1201.5197}.

\bibitem{Kowal_Reconnection_HardSpec}
G.~Kowal, E.~M. de~Gouveia Dal~Pino and A.~Lazarian, \emph{Magnetohydrodynamic
  simulations of reconnection and particle acceleration: Three-dimensional
  effects}, {\emph{The Astrophysical Journal 735 (July, 2011) 102} (2011)
  arXiv:1103.2984}.

\bibitem{Drury_1stOFMReconnection_HardSpec}
G.~Kowal, E.~M. de~Gouveia Dal~Pino and A.~Lazarian, \emph{First-order fermi
  acceleration driven by magnetic reconnection}, {\emph{Monthly Notices of the
  Royal Astronomical Society 422 (May, 2012) 2474?2476} (2012)
  arXiv:1201.6612}.

\bibitem{Alves_TidalDisruption_HardSpec}
R.~Alves~Batista and J.~Silk, \emph{Ultrahigh-energy cosmic rays from
  tidally-ignited white dwarfs}, {\emph{Physical Review D 96 (Nov., 2017)
  103003} (2017) arXiv:1702.06978}.

\bibitem{Bihel_TidalDisruption_HardSpec}
D.~Biehl, D.~Boncioli, C.~Lunardini and W.~Winter, \emph{Tidally disrupted
  stars as a possible origin of both cosmic rays and neutrinos at the highest
  energies}, {\emph{Sci. Rep. 8 (2018) 1082} (2018) arXiv:1711.03555}.

\bibitem{Zhang_TidalDisruption_HardSpec}
B.~T. Zhang, K.~Murase, L.~Oikonomou and Z.~Li, \emph{High-energy cosmic ray
  nuclei from tidal disruption events: Origin, survival and implications},
  {\emph{Phys. Rev. D96 (2017) 063007} (2018) arXiv:1706.00391}.

\bibitem{Guepin_TidalDisruption_HardSpec}
C.~Gu\'epin, K.~Kotera, E.~Barausse, K.~Fang and K.~Murase, \emph{Ultra-high
  energy cosmic rays and neutrinos from tidal disruptions by massive black
  holes}, {\emph{Astron. Astrophys. 616 (2018) A179} (2018) arXiv:1711.11274}.

\bibitem{the_pierre_auger_collaboration_combined_2017}
T.~P. Collaboration, \emph{Combined fit of spectrum and composition data as
  measured by the {Pierre Auger} observatory},
  \href{https://arxiv.org/abs/1612.07155}{{\ttfamily 1612.07155}}.

\bibitem{batista_cosmogenic_2019}
R.~A. Batista, R.~M. de~Almeida, B.~Lago and K.~Kotera, \emph{Cosmogenic photon
  and neutrino fluxes in the {Auger} era},
  \href{https://doi.org/10.1088/1475-7516/2019/01/002}{\emph{JCAP} {\bfseries
  2019} (2019) 002} [\href{https://arxiv.org/abs/1806.10879}{{\ttfamily
  1806.10879}}].

\bibitem{evans_thechandra_2010}
I.~N. Evans and {others}, \emph{{The CHANDRA} {Source} {Catalog}},
  \href{https://doi.org/10.1088/0067-0049/189/1/37}{\emph{The Astrophysical
  Journal Supplement Series} {\bfseries 189} (2010) 37}.

\bibitem{van_vliet_determining_2019}
A.~van Vliet, R.~A. Batista and J.~R. Hörandel, \emph{Determining the fraction
  of cosmic-ray protons at ultra-high energies with cosmogenic neutrinos},
  {\emph{{arXiv}:1901.01899 [astro-ph]} (2019) }
  [\href{https://arxiv.org/abs/1901.01899}{{\ttfamily 1901.01899}}].

\end{thebibliography}\endgroup

\end{document}